\documentclass[pre,twocolumn,superscriptaddress]{revtex4}

\usepackage{graphicx}
\usepackage{dcolumn}
\usepackage{amsmath}
\usepackage{algorithm}
\usepackage{algpseudocode}
\usepackage{enumitem}
\usepackage{subcaption}
\usepackage{hyperref}

\newcommand{\etal}{\textit{et~al.}}
\newcommand{\e}{\mathrm{e}}
\newcommand{\Ord}{\mathrm{O}}

\begin{document}
\title{Hierarchical core-periphery structure in networks}

\author{Austin Polanco}
\affiliation{Department of Physics, University of Michigan, Ann Arbor, Michigan 48109, USA}

\author{M. E. J. Newman}
\affiliation{Department of Physics, University of Michigan, Ann Arbor, Michigan 48109, USA}
\affiliation{Center for the Study of Complex Systems, University of Michigan, Ann Arbor, Michigan 48109, USA}

\begin{abstract}
  We study core-periphery structure in networks using inference methods based on a flexible network model that allows for traditional onion-like cores within cores, but also for hierarchical tree-like structures and more general non-nested types of structure.  We propose an efficient Monte Carlo scheme for fitting the model to observed networks and report results for a selection of real-world data sets.  Among other things, we observe an empirical distinction between networks showing traditional core-periphery structure with a dense core weakly connected to a sparse periphery, and an alternative structure in which the core is strongly connected both within itself and to the periphery.  Networks vary in whether they are better represented by one type of structure or the other.  We also observe structures that are a hybrid between core-periphery structure and community structure, in which networks have a set of non-overlapping cores that correspond roughly to communities, surrounded by a single undifferentiated periphery.  Computer code implementing our methods is available.
\end{abstract}
\maketitle

\section{Introduction}
Networks are widely used as a compact and convenient mathematical representation of the connections between the elements of a complex system, such as data connections on the Internet, citations between papers, social contacts among people or animals, synaptic connections between brain cells, and biological and biochemical networks of many kinds~\cite{Boccaletti06,Newman18c}.  A significant amount of effort has been devoted in recent years to analyzing and understanding large-scale structure in such networks, especially community structure~\cite{GN02,Fortunato10}, but also nested~\cite{GN02,CMN08} and overlapping~\cite{PDFV05,ABFX08} communities and stratification~\cite{HRH02,NP15}, as well as the related issues of embedding and graph representation learning~\cite{HYL17}.  In this paper we focus on a less well-studied form of large-scale structure, \textit{core-periphery structure}, in which a network is divided into a densely connected core of nodes surrounded by a sparser periphery~\cite{BE99,Holme05b,RPFM14}.  The observation of core-periphery structure communicates different information about a network from community structure.  Where community structure deals with the identification of groups or types of nodes, core-periphery structure focuses on their roles and centrality.  Core-periphery structure is integral to understanding the link between node position and function in networks, for instance in the Internet~\cite{Carmi07,ADBV08}, neuroscience~\cite{Bassett13}, and economics~\cite{FRS10}.

\begin{figure*}[t]
    \begin{subfigure}{4.5cm}
        \includegraphics[width=\textwidth]{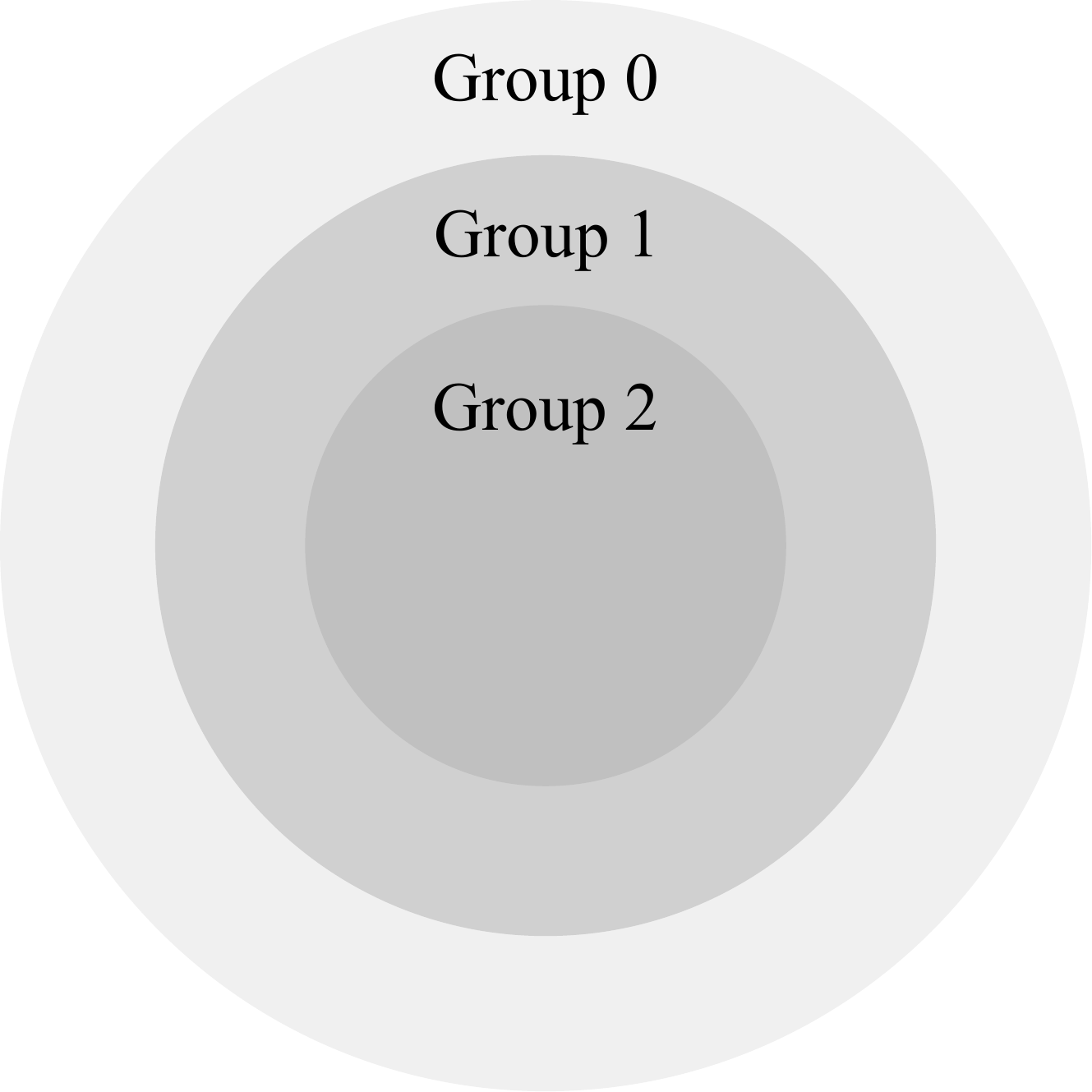}
        \caption{Nested groups}
    \end{subfigure}
    \hfill
    \begin{subfigure}{6.5cm}
        \includegraphics[width=\textwidth]{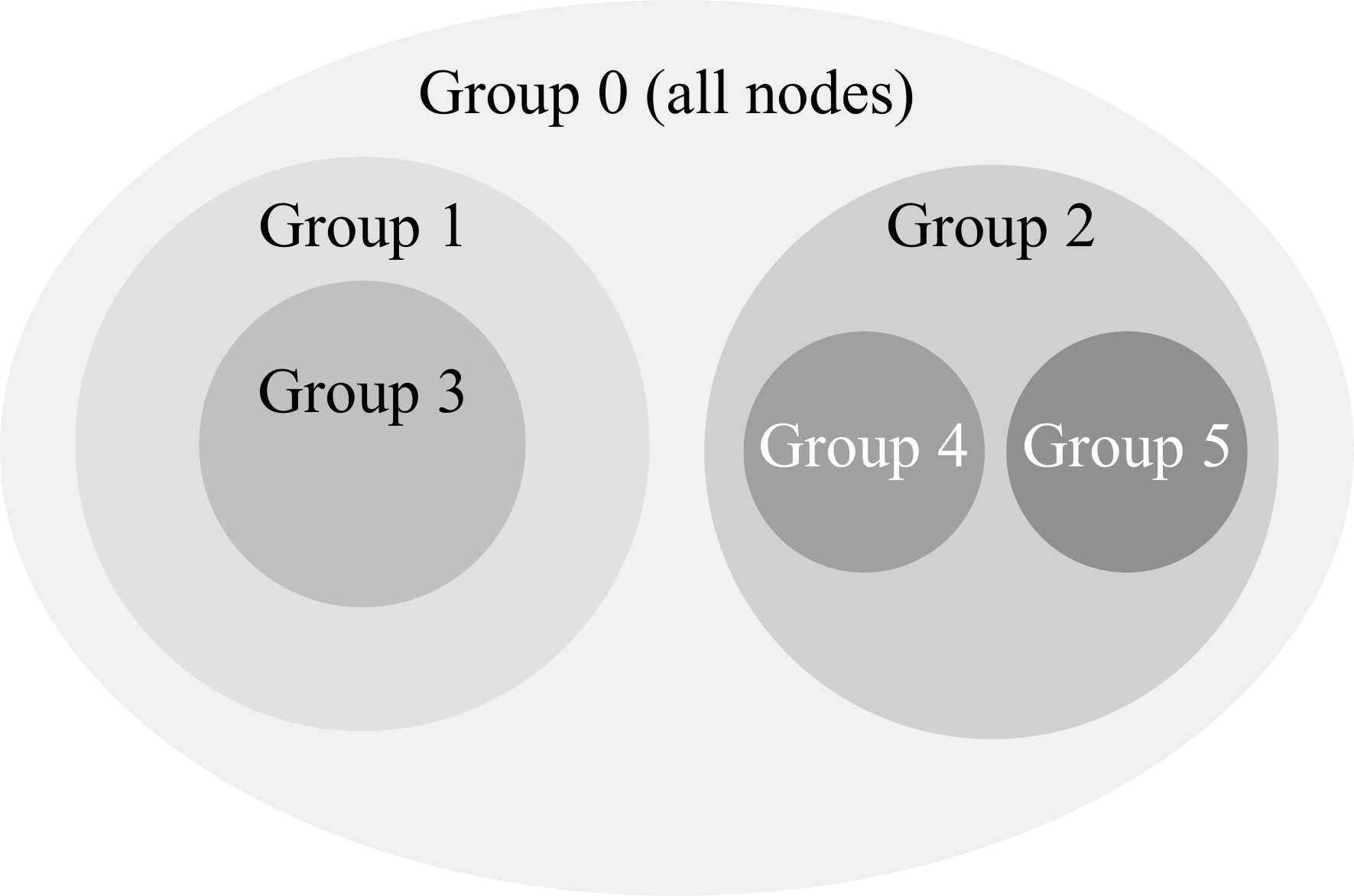}
        \caption{Tree-like hierarchy}
    \end{subfigure}
    \hfill
    \begin{subfigure}{6.5cm}
        \includegraphics[width=\textwidth]{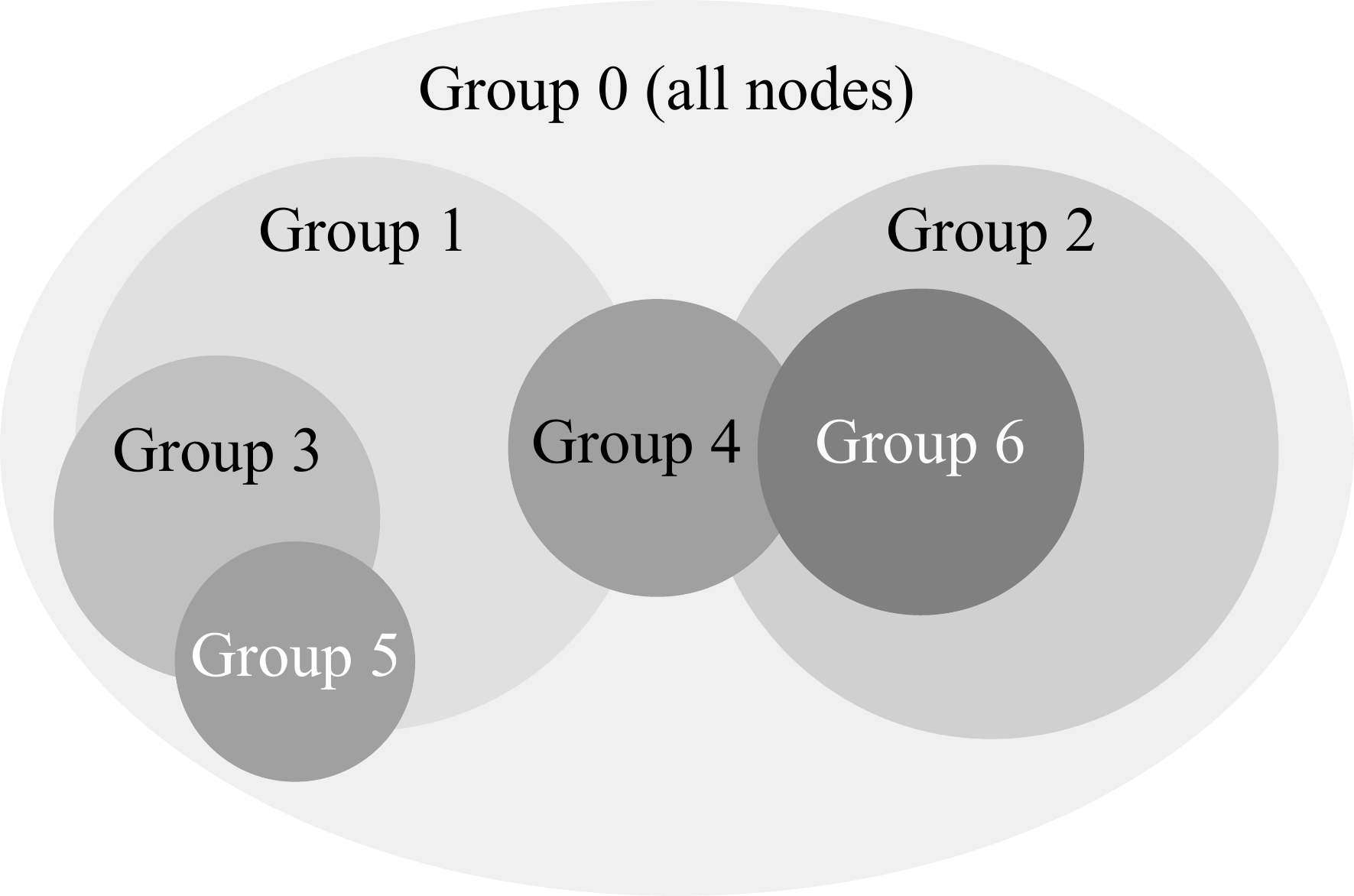}
        \caption{General core-periphery structure}
    \end{subfigure}
    \caption{The model proposed here is capable of capturing a range of different types of structure, including (a)~traditional nested groups, (b)~tree-like branching structures, and (c)~complex structures of overlapping groups.}
    \label{fig:schematic}
\end{figure*}

A range of heuristic methods have been proposed in the past for detecting core-periphery structure in networks.  In the simplest case the challenge is to take unlabelled network data and assign each node of the network in some automated fashion to either the core or the periphery.  In more complex cases one may attempt to infer a onion-like sequence of deeper and deeper cores within cores.

The problem, however, is not entirely well posed, since we have not precisely defined what ``core'' and ``periphery'' mean.  Various researchers have chosen to define them in various ways and there is, as a result, a corresponding spectrum of algorithmic approaches.  Perhaps the oldest approach is the \textit{$k$-core decomposition}, in which nodes are recursively removed from a network in order of increasing degree and the sequence in which they are removed defines a sliding scale from core to periphery~\cite{ADBV06}.  This method essentially equates the core with high-degree nodes.  Another well-established method is that of Borgatti and Everett~\cite{BE99}, who defined a quality function akin to the well-known modularity function for community detection~\cite{NG04}.  Borgatti and Everett's function takes as input a network and a putative division into core and periphery and returns a score that indicates whether the division is a ``good'' one in a certain sense.  Then by maximizing the function over all possible divisions one can find the ``best'' core-periphery decomposition.  Variants of the same idea have been explored by Rombach~\etal~\cite{RPFM14}, as well as by Kojaku and Masuda~\cite{KM17} who proposed a multi-group version.  Cucuringu~\etal~\cite{CRLP16} have explored several methods for finding core-periphery structure based on counting geodesic paths and on spectral network properties.  Other methods use node-level properties of the network such as the clustering coefficient~\cite{Holme05b} or centrality measures~\cite{LCP14}.

Stochastic block models, commonly used in community detection~\cite{HLL83,KN11a}, have also been applied to core-periphery structure.  In these models, one assumes that nodes are divided into types (e.g.,~core and periphery) and that the probability~$\omega_{rs}$ of an edge between a pair of nodes depends on the types~$r$ and~$s$ of the nodes.  Zhang~\etal~\cite{ZMN15} studied a two-group version of this model with $\omega_{11} > \omega_{12} > \omega_{22}$, which generates networks with classic core-periphery structure, then fitted the model to empirical network data to detect structure.  Gallagher~\etal~\cite{GYF21} took a similar approach but with more than two groups and multiple nested cores such that $\omega_{rs} = f[\max(r,s)]$ for some function~$f[r]$.

In this paper we also take a model-fitting approach to the study of core-periphery structure, but we formulate a more general model that includes the classic two-group structure but also allows for more flexible structures as well.  In particular, we consider a hierarchical model with any number of groups, the number being determined by the network structure using Bayesian model selection, and we specifically allow for the possibility that the hierarchy may not be perfectly nested and moreover that there maybe multiple cores and multiple peripheries at any level in the hierarchy.  These features allow our model to capture variants of core-periphery structure that occur in real-world networks but are not captured by traditional analyses.  In a multilevel hierarchy, for instance, an inner core may not be perfectly contained within an outer one.  In a network with community structure the individual communities may each have a separate core of their own, or there may be several cores within a single periphery.  By fitting our proposed model to network data we show how it can reveal subtle structure and, in the process, encounter some new formations within networks that have not previously received significant attention.


\section{A hierarchical model of\\network structure}
One can think of conventional core-periphery structure as a hierarchy.  In the simplest case the hierarchy has just two levels, an outer periphery and an inner core.  In more elaborate cases there can be an onion-like succession of levels, each one nested inside the next, from the outermost to the innermost---see Fig.~\ref{fig:schematic}a.  This, however, is not the most general kind of nested structure.  A more general structure can have any number of first-level cores within a single periphery, and any number of second-level cores within those first-level ones, and so forth (Fig.~\ref{fig:schematic}b).  A structure like this can be captured with a tree or dendrogram, and some attempts have been made to fit dendrograms to networks~\cite{CMN08}.  The work presented here goes still further by allowing for the possibility that the hierarchy is not perfectly nested.  In our model, we envisage a hierarchical series of cores, but we allow these cores to be placed in any way around the network, with no requirement that they be nested within one another (Fig.~\ref{fig:schematic}c).  In practice, if two cores do not overlap then they do not interact and hence their relative order in the hierarchy has no effect, but if they overlap then the higher ranked one takes precedence over the lower in a manner we will describe.  The net result is a hierarchical model that generalizes both the conventional onion layers and the dendrogram and, as we will see, captures a wide range of possible structures.  We then fit this model to network data to infer core-periphery structure in real-world networks.  In detail the procedure is as follows.

Consider an unweighted, undirected network of $n$ nodes, which can belong to any of $k$ groups labeled by $r=0\ldots k-1$.  By contrast with traditional community structure models we allow nodes to belong to any number of groups simultaneously, up to and including all of them.  In addition, every node always belongs to group~0, which acts as a sort of default or base group.  Thus each node belongs to group~0 plus some selection of the other groups $1\ldots k-1$.  (We label the first group~0 rather than~1 to remind ourselves of its special status.)

We define a set of indicator variables $g_u^r$ such that $g_u^r=1$ if node~$u$ belongs to group~$r$ and 0 otherwise.  Then we define a set of probabilities~$\omega_r$, one for each group including group~0, and place edges between node pairs independently with probability~$\omega_r$, where $r$ is the \emph{highest common group} that both nodes of the pair belong to, meaning the one with the highest number.  For example, if one node belongs to groups 0, 1, 2 and another belongs to 0, 1, 3, then there is an edge between them with probability~$\omega_1$.

Figure~\ref{fig:schematic}c illustrates the behavior of the resulting network: the groups form ``patches'' that lie one on top of another and the topmost visible patch takes precedence for each node pair.  For instance, if all nodes belong to group~0 only, then every pair of nodes has equal probability~$\omega_0$ of being connected, which gives us a standard random graph.  But if we assign some subset of the nodes to group~1 then for any pair of nodes that are both in group~1 the probability of an edge becomes~$\omega_1$ and overrides the previous~$\omega_0$.  Similarly any pair of nodes assigned to group~2, including nodes already in groups 0 and~1, have probability~$\omega_2$ of connection, overriding $\omega_0$ and~$\omega_1$, and so forth.  The end result is a model that starts out as a random graph but then adds variation and detail to the network wherever it is needed to capture local structural features.  This approach has some similarities to those of Kojaku and Masuda~\cite{KM17} and Gallagher~\etal~\cite{GYF21}, but it differs crucially in that it does not force the groups to be either strictly nested within each other or nonoverlapping.  Given the definition of the model, our goal is now to fit it to observed network data to infer the best choice of groups~$g_u^r$ for each node.

Suppose we have a network of $n$ nodes, represented by its $n\times n$ adjacency matrix~$A$ with elements $a_{uv} = 1$ if there is an edge between nodes~$u$ and~$v$ and 0 otherwise.  Then the probability of observing a particular network if it was generated from our model with given~$k$ and given group memberships~$g$ is
\begin{align}
  P(A|\omega,k,g) &= \prod\limits_{u<v}\omega_{h(u,v)}^{a_{uv}}
  \bigl[ 1-\omega_{h(u,v)} \bigr]^{1-a_{uv}} \nonumber\\
  &=\prod_{r=0}^{k-1}\omega_r^{m_r}(1-\omega_r)^{t_r - m_r},
\end{align}
where $h(u,v)$ is the highest common group of nodes $u$ and~$v$, and
\begin{equation}
t_r = \sum_{u<v} \delta_{r,h(u,v)}
\end{equation}
is the number of node pairs with highest common group~$r$, and
\begin{equation}
m_r = \sum_{u<v} a_{uv} \delta_{r,h(u,v)}
\end{equation}
is the number of such pairs that are connected by an edge.

Our primary interest in performing the fit is to determine the group memberships~$g_u^r$.  The values of the parameters~$\omega_r$ are not of particular interest, so we eliminate them by marginalizing.  We assume a uniform prior~$P(\omega_r)=1$ for all~$r$ and write
\begin{align}
P(A|k,g) &= \int P(A,\omega|k, g)\>d\omega \nonumber\\
  &= \int P(A|\omega, k, g)P(\omega)\>d\omega \nonumber\\
  &= \prod_{r=0}^{k-1}\,\int_0^1\omega_r^{m_r}(1-\omega_r)^{t_r - m_r}
     \>d\omega_r \nonumber\\
  &= \prod_{r=0}^{k-1} \frac{m_r!(t_r-m_r)!}{(t_r+1)!}.
\end{align}

\subsection{Prior on group assignments}
We consider two different scenarios.  In the first, the number~$k$ of groups in the network is fixed; in the second it is not.  The number might be fixed if, for instance, our goal is to find traditional core-periphery structure, for which there are always two groups, the core and the periphery.  In other cases, we may be interested in allowing the number of groups to vary and determining what number best fits the network we observe.

If the number of groups is fixed, we can write
\begin{equation}
P(g|A,k) = \frac{P(A|g,k)P(g|k)}{P(A|k)},
\label{eq:PgAk}
\end{equation}
and we can maximize this quantity to find the most probable values of~$g$, or sample from the distribution it defines to generate plausible core-periphery structures in proportion to their probability.  In this paper we do the latter, using a Monte Carlo method.

To do this we need to make a choice for the prior~$P(g|k)$.  Naively we might assume a prior that is uniform over all assignments~$g$, but this would be a mistake.  Such a choice produces group sizes that are narrowly peaked about $\frac12 n$, which is highly unrealistic.  Similar problems occur in traditional community detection~\cite{Peixoto17,RCRN17}, where a good solution is instead to choose the \emph{sizes} of the groups to be uniform rather than the assignments and we take an analogous approach here.  For each group~$r>0$ we first choose a size~$n_r$ uniformly at random between 0 and~$n$, each value thus having probability $1/(n+1)$.  Then we choose uniformly at random one of the ${n\choose n_r}$ ways to assign $n_r$ nodes to the group, each choice thus having probability $1/{n\choose n_r}$.  Hence for all groups the total probability of an assignment is
\begin{align}
P(g|k) = \prod_{r=1}^{k-1} \frac{1}{(n+1){n\choose n_r}}
  = \prod_{r=1}^{k-1} \frac{n_r!(n-n_r)!}{(n+1)!}.
\label{eq:Pgk}
\end{align}

\subsection{Prior on number of groups}
\label{section:prior_ngroups}
In traditional core-periphery structure calculations one considers the existence of a single core and a single periphery, and this is the approach taken for instance in~\cite{ZMN15}.  If our goal, however, is to find multiple groups of unknown number, including multiple or overlapping cores, then we need to allow~$k$ to vary, which means choosing a prior on~$k$.  Here we adopt the approach taken in~\cite{MMFH13,NR16} and use a Poisson distribution with mean~1:
\begin{align}
P(k) = {\e^{-1}\over (k-1)!}.
\label{eq:Pk}
\end{align}
Note that group~0 always exists, so the distribution of the number of groups is effectively a distribution over $k-1$, which is why we have $1/(k-1)!$ in the denominator.

With this choice, we can now write
\begin{align}
P(g,k|A) &= P(k) P(g|A,k)
  = {P(k) P(g|k) P(A|g,k)\over P(A)} \nonumber\\
  &\propto {1\over (k-1)!} \prod_{r=1}^{k-1} {n_r!(n-n_r)!\over(n+1)!} \prod_{r=0}^{k-1}
   {m_r! (t_r-m_r)!\over(t_r+1)!}.
\label{eq:full_post}
\end{align}
Again, we can sample from this probability to generate a selection of values~$k$ and group assignments~$g$ that are representative of the network.  In the following section we describe the algorithm we use to achieve this.

\section{Sampling from the posterior distribution}
We sample from the posterior distribution~\eqref{eq:full_post} using a Markov-chain Monte Carlo method.  We describe the method first for the simpler case where the number of groups~$k$ is fixed, then for the more complicated case of varying~$k$.

\subsection{Monte Carlo algorithm for the case of fixed $k$}
For the case of fixed~$k$ our Monte Carlo scheme is as follows.
\begin{enumerate}
\item We choose a group $s$ uniformly at random from $1\ldots k-1$.
\item With equal probability~$\frac12$ we propose to either remove a node from group~$s$ or add a node to it.  If we are removing, the node to be removed is chosen uniformly at random from those currently in the group.  If there are no nodes in the group, we do nothing and move on to the next Monte Carlo step.  If we are adding, the node to be added is chosen uniformly at random from those currently not in the group.  If the group is full---all $n$ nodes are already members---we do nothing and move on to the next step.
\item The proposed move is accepted with the Metropolis-Hastings style acceptance probability
\begin{equation}
\alpha(g\to g') = \min\biggl(1,{P(A|g',k)\over P(A|g,k)}\biggr),
\label{eq:pa}
\end{equation}
where $g'$ represents the group assignments after the addition or removal.  If the move is accepted, the chosen node is added or removed as proposed.  If the move is not accepted, the group assignments $g$ remain unchanged on this step.
\item Repeat from step~1.
\end{enumerate}

In the limit where this algorithm tends to an equilibrium distribution of states, that distribution will be the one given in Eq.~\eqref{eq:PgAk}.  To demonstrate this, it suffices to prove two results: first that the algorithm is ergodic and second that it satisfies detailed balance.  Ergodicity requires that every state of the system be accessible from every other by a finite sequence of moves.  This is trivially true in the present case, since the membership of any group can be set to anything we like in at most $n$ moves by first removing any nodes we don't want and then adding in those we do.

Detailed balance is a little more complicated.  Detailed balance requires that in equilibrium the average rate of moves $g\to g'$ equals the average rate $g'\to g$, which means
\begin{equation}
P(g|A,k) P(g\to g') = P(g'|A,k) P(g'\to g),
\label{eq:db1}
\end{equation}
where $P(g\to g')$ is the probability of making the transition $g\to g'$.  This probability can be written as
\begin{equation}
P(g\to g') = \pi(g\to g') \alpha(g\to g'),
\end{equation}
where $\pi(g\to g')$ is the probability of proposing the move and $\alpha(g\to g')$ is the probability of accepting it as in Eq.~\eqref{eq:pa}.  Then Eq.~\eqref{eq:db1} can be written as
\begin{equation}
{P(g'|A,k)\over P(g|A,k)}
  = {\pi(g\to g')\,\alpha(g\to g')\over\pi(g'\to g)\,\alpha(g'\to g)}.
\label{eq:db2}
\end{equation}
We can show that this condition is satisfied by the proposed Monte Carlo algorithm as follows.

From Eq.~\eqref{eq:PgAk} we have
\begin{align}
{P(g'|A,k)\over P(g|A,k)} = {P(A|g',k)P(g'|k)\over P(A|g,k)P(g|k)},
\label{eq:db3}
\end{align}
while from Eq.~\eqref{eq:pa} the ratio of the two acceptance probabilities is
\begin{equation}
{\alpha(g\to g')\over\alpha(g'\to g)} = {P(A|g',k)\over P(A|g,k)}.
\label{eq:alpharatio}
\end{equation}
Substituting~\eqref{eq:db3} and~\eqref{eq:alpharatio} into~\eqref{eq:db2}, a factor of $P(A|g',k)/P(A|g,k)$ cancels and we are left with
\begin{equation}
{P(g'|k)\over P(g|k)} = {\pi(g\to g')\over\pi(g'\to g)}.
\label{eq:db5}
\end{equation}
If our Monte Carlo algorithm satisfies this condition, then it satisfies detailed balance.

Using Eq.~\eqref{eq:Pgk}, the left-hand side can be written as
\begin{align}
{P(g'|k)\over P(g|k)} = \prod_{r=1}^{k-1} {n_r'!(n-n_r')!\over n_r!(n-n_r)!}\,,
\label{eq:db4}
\end{align}
where $n_r$ is the number of nodes in group~$r$ before the move and $n_r'$ is the number afterwards.  Suppose the particular move we are considering $g\to g'$ is one that adds a node to group~$s$.  Then $n_s'=n_s+1$, while $n_r'=n_r$ for all other groups, so~\eqref{eq:db4} simplifies to
\begin{align}
{P(g'|k)\over P(g|k)} = {(n_s+1)!(n-n_s-1)!\over n_s!(n-n_s)!}
  = {n_s+1\over n-n_s}.
\label{eq:dblhs}
\end{align}

For the right-hand side of Eq.~\eqref{eq:db5}, for the same move that adds a node to group~$s$, the proposal probability is
\begin{equation}
\pi(g\to g') = {1\over k-1} \times \frac12 \times {1\over n-n_s}
  = {1\over 2(k-1)(n-n_s)}.
\end{equation}
Here the factor $1/(k-1)$ is the probability of choosing the particular group~$r$ out of all $k-1$ possibilities, the factor $\frac12$ is the probability of choosing to add a node, and the factor $1/(n-n_s)$ is the probability of choosing the particular node to be added from the $n-n_s$ possibilities.  Meanwhile, for the reverse move $g'\to g$, which involves removing the same node from group~$s$ again, the proposal probability is
\begin{equation}
\pi(g'\to g) = {1\over k-1} \times \frac12 \times {1\over n_s+1}
  = {1\over 2(k-1)(n_s+1)},
\end{equation}
since there are now $n_s+1$ nodes in the group.  The ratio of the two proposal probabilities is thus
\begin{equation}
{\pi(g\to g')\over\pi(g'\to g)}
  = {1/2(k-1)(n-n_s)\over1/2(k-1)(n_s+1)} = {n_s+1\over n-n_s},
\end{equation}
which agrees with Eq.~\eqref{eq:dblhs} and hence Eq.~\eqref{eq:db5} is satisfied and detailed balance is obeyed in this instance.  The proof for the case where we remove a node from group~$s$ follows the same lines and leads to the same conclusion: the algorithm satisfies detailed balance and hence samples correctly from the target distribution $P(g|A,k)$ in equilibrium.

\subsection{Algorithm for varying $k$}
When $k$ is allowed to vary the algorithm is more complex, involving two types of moves that each take us from a combined state~$(g,k)$ to a state~$(g',k')$, as follows.

\medskip
\textbf{Type 1}: In a move of type~1 we choose a group~$s$ uniformly at random from $1\ldots k-1$.  With probability $\frac12$ we add a new node to the group chosen uniformly from the set of nodes that do not currently belong; otherwise, we remove an existing node from the group, chosen uniformly from those in the group.  If we choose to add a node but group~$s$ is already full then we do nothing and move on to the next Monte Carlo step.  If we choose to remove a node and group~$s$ is already empty then the entire group is deleted and the number of groups~$k$ decreases by one, with the labels of all groups above $s$ also decreasing by one so that they still run to a maximum of~$k-1$.

\textbf{Type 2}: In a move of type~2 we choose a group index~$s$ uniformly at random from $1\ldots k$.  We increase by one the labels of all groups~$s$ and greater (if there are any), create a new empty group with label~$s$, and increase the value of~$k$ by one.

\medskip With these definitions, the complete algorithm is now as follows:
\begin{enumerate}
  \item With probability $1-1/2k(n+1)$ propose a move of type~1.
    \begin{enumerate}[label=\theenumi\alph*)]
    \item If $k=1$ do nothing, since this implies all nodes are in group~0 only, so there are no moves to be made and there is no change of state on this Monte Carlo step.
    \item Otherwise when $k>1$ choose a random move of type~1.
   \end{enumerate}
  \item Else, with probability $1/2k(n+1)$, choose a random move of type~2.
  \item Accept the proposed move with probability
\begin{align}
\alpha(g,k\to g,k) = \min \biggl( 1,\frac{P(A|g',k')}{P(A|g,k)} \biggr).
\label{eq:a_rat}
\end{align}
Accepted moves are performed as proposed.  If the move is not accepted the state of the system remains unchanged.
  \item Repeat from step~1.
\end{enumerate}

    


    




        

      

This algorithm again satisfies the condition of ergodicity trivially: we can reach any state with any number of groups in a finite number of moves by first removing all nodes from all groups except group~0, then removing the groups themselves, then adding back the appropriate number of groups and filling them with the desired nodes.  The algorithm also satisfies the condition of detailed balance, which for this algorithm takes the form
\begin{equation}
  \frac{P(g',k'|A)}{P(g,k|A)} = \frac{\pi(g,k\to g',k')\,\alpha(g,k\to g',k')}{\pi(g',k'\to g,k)\,\alpha(g',k'\to g,k)}.
\label{eq:dbfull1}
\end{equation}
The left-hand side can be written as
\begin{equation}
\frac{P(g', k'|A)}{P(g, k|A)} = {P(g',k')P(A|g',k')\over P(g,k) P(A|g,k)},
\label{eq:dbfull2}
\end{equation}
and the ratio of acceptance probabilities is
\begin{equation}
{\alpha(g,k\to g',k')\over\alpha(g',k'\to g,k)} = {P(A|g',k')\over P(A|g,k)}.
\label{eq:dbfull3}
\end{equation}
Substituting from~\eqref{eq:dbfull2} and~\eqref{eq:dbfull3} into~\eqref{eq:dbfull1}, a factor of $P(A|g',k')/P(A|g,k)$ cancels and our detailed balance condition reduces to
\begin{equation}
{P(g',k')\over P(g,k)} = \frac{\pi(g,k\to g',k')}{\pi(g',k'\to g,k)}.
\label{eq:dbsimple1}
\end{equation}

From Eqs.~\eqref{eq:Pgk} and~\eqref{eq:Pk} the left-hand side is
\begin{align}
\frac{P(g', k')}{P(g, k)} = \frac{(k-1)!\prod_{r=1}^{k-1} n'_r!(n-n'_r)!/(n+1)!}%
  {(k'-1)!\prod_{r=1}^{k'-1} n_r!(n-n_r)!/(n+1)!}.
\label{eq:prob_t1t1}
\end{align}
Consider first the case where we propose a move of type~1 that adds a node to group~$s$.  Then $k'=k$ and $n_s' = n_s+1$, and $n_r'=n_r$ for all other groups~$r$, so Eq.~\eqref{eq:prob_t1t1} becomes
\begin{equation}
\frac{P(g', k')}{P(g, k)} = {n_s+1\over n-n_s}.
\label{eq:dbsimple2}
\end{equation}
The probability of proposing such a move is
\begin{align}
\pi&(g,k\to g',k') \nonumber\\
  &= \biggl( 1 - {1\over2k(n+1)} \biggr) \times {1\over k-1}
  \times \frac12 \times {1\over n-n_s} \nonumber\\
  &= \biggl( 1 - {1\over2k(n+1)} \biggr) {1\over2(k-1)(n-n_s)},
\end{align}
while the probability of proposing the reverse move is
\begin{equation}
\pi(g',k'\to g,k) 
  = \biggl( 1 - {1\over2k(n+1)} \biggr) {1\over2(k-1)(n_s+1)}.
\end{equation}
Thus the ratio of the two is
\begin{equation}
{\pi(g,k\to g',k')\over \pi(g',k'\to g,k)} = {n_s+1\over n-n_s}.
\label{eq:dbsimple3}
\end{equation}
Between Eqs.~\eqref{eq:dbsimple2} and~\eqref{eq:dbsimple3}, our detailed balance condition~\eqref{eq:dbsimple1} is now satisfied.  By a similar argument we can show that detailed balance is also satisfied when a node is removed from a group.

Now consider a move of type~2, which creates a new empty group with a random label~$s$.  For such a move Eq.~\eqref{eq:prob_t1t1} becomes 
\begin{align}
\frac{P(g',k')}{P(g, k)} &= \frac{(k-1)!\prod_{r=1}^k n'_r!(n-n'_r)!/(n+1)!}{k!\prod_{r=1}^{k-1} n_r!(n-n_r)!/(n+1)!} \nonumber\\
  &= \frac{1}{k(n+1)}\,,
\label{eq:dbfull4}
\end{align}
where all factors inside the products have canceled except for those pertaining to the new group, which gives us the factor of $1/(n+1)$.

The proposal probability for this move is equal to the probability that we decide to do a move of type~2 times the probability that we choose to add a new group with a particular label~$s$ out of the $k$ possibilities, giving
\begin{align}
\pi(g,k\to g',k') = {1\over2k(n+1)} \times {1\over k}
  = {1\over2k^2(n+1)}.
\end{align}
The reverse move on the other hand occurs when we perform a move of type~1 and choose group~$s$ from the $k$ possibilities, then attempt to remove a node only to discover that the group is already empty, causing us to delete the entire group.  The proposal probability for this move is
\begin{align}
\pi(g',k'\to g,k) &= \biggl( 1 - {1\over2(k+1)(n+1)} \biggr) \times {1\over k}
  \times {1\over2} \nonumber\\
  &= {1\over2k} \biggl( 1 - {1\over2(k+1)(n+1)} \biggr).
\end{align}
Now the ratio of the two probabilities is
\begin{align}
{\pi(g,k\to g',k')\over\pi(g',k'\to g,k)}
  &= {4k(k+1)(n+1)/(2(k+1)(n+1)-1)\over 2k^2(n+1)} \nonumber\\
  &= {2(k+1)/k\over2(k+1)(n+1)-1} \nonumber\\
  &= {1\over k(n+1)} + \Ord(1/n^2).
\label{eq:dbfull5}
\end{align}
Here we assume $n$ is large and hence that terms of order~$1/n^2$ can be neglected, making Eq.~\eqref{eq:dbfull5} equal to Eq.~\eqref{eq:dbfull4}, and hence our detailed balance condition, Eq.~\eqref{eq:dbsimple1}, is satisfied.

This completes the proof of correctness of our algorithms.  In the following sections we apply these algorithms to fit our model to a variety of networks in order to study core-periphery structure.

\begin{figure*}[ht]
  \begin{subfigure}[t]{\columnwidth}
      \includegraphics[width=\textwidth]{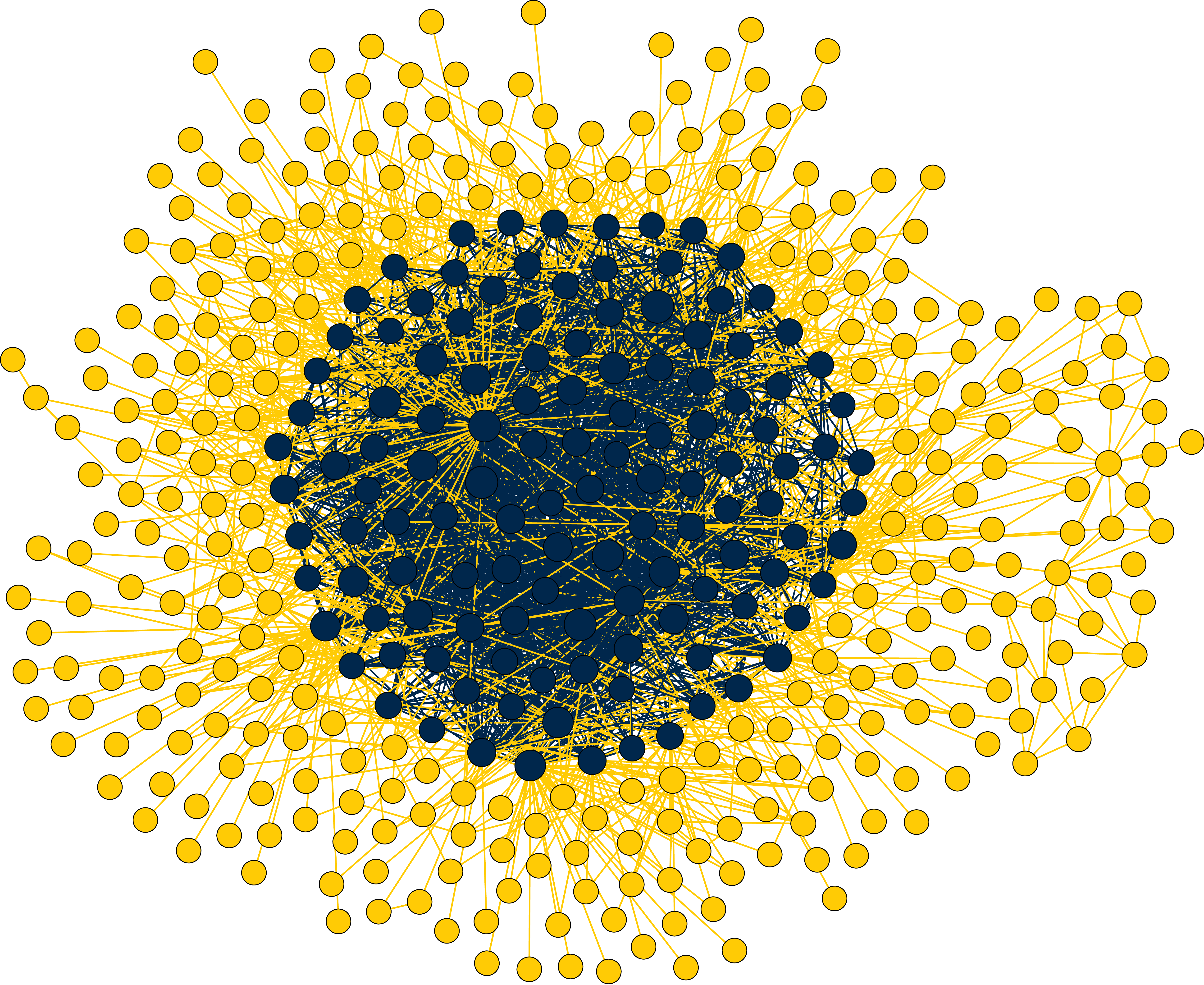}
      \caption{Airline routes among European airports~\cite{CGZ13}}
      \label{fig:euair}
  \end{subfigure}
  \hfill
  \begin{subfigure}[t]{1\columnwidth}
      \includegraphics[width=\textwidth]{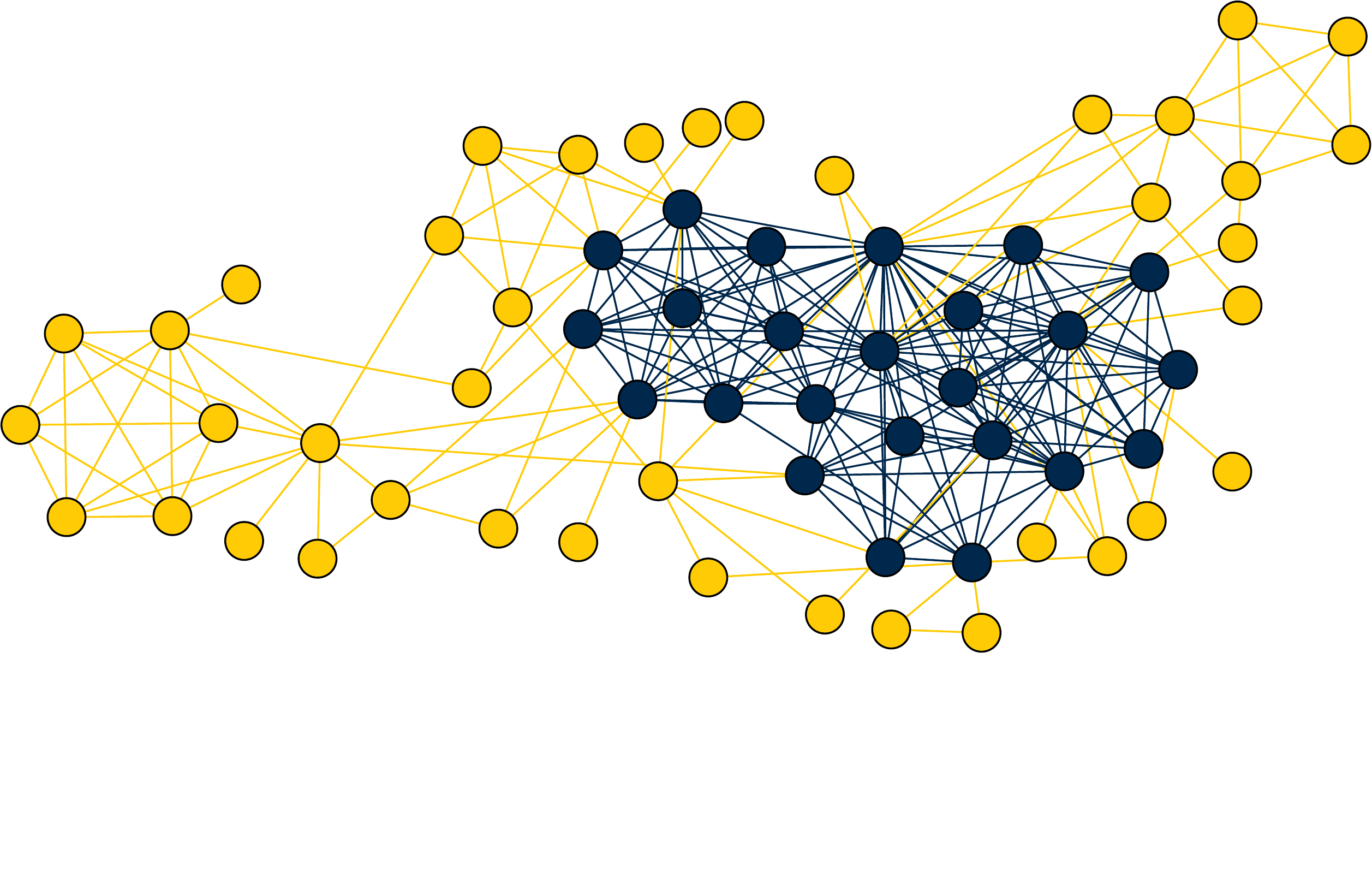}
      \caption{A network of associations among terrorists involved in the 2004 Madrid train bombing~\cite{Hayes06}}
      \label{fig:mad_terr}
  \end{subfigure}
  \begin{subfigure}[t]{1\columnwidth}
      \includegraphics[width=\textwidth]{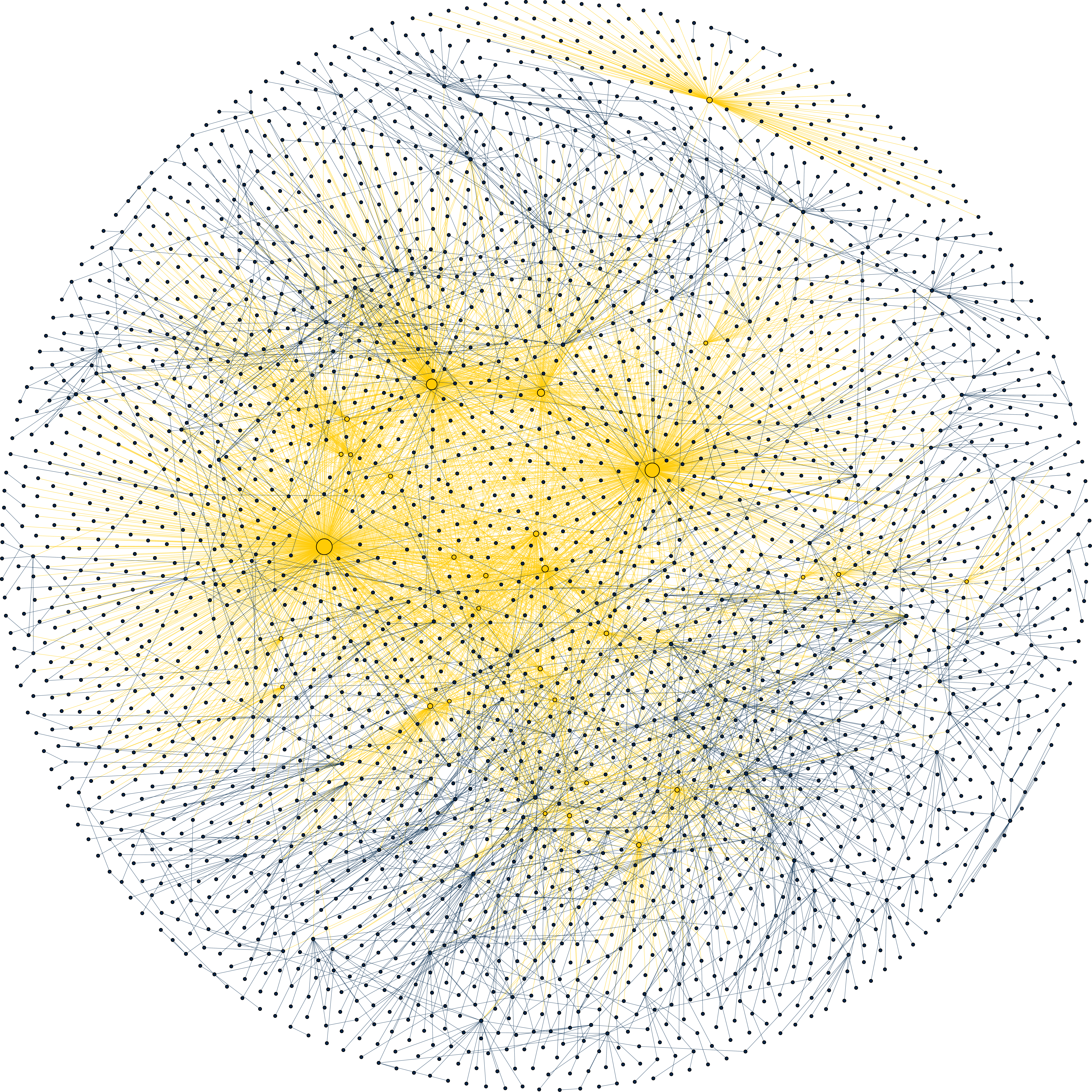}
      \caption{Network representation of the Internet in November 1997 at the autonomous system level~\cite{LA14}}
      \label{fig:internet}
  \end{subfigure}
  \hfill
  \begin{subfigure}[t]{1\columnwidth}
      \includegraphics[width=\textwidth]{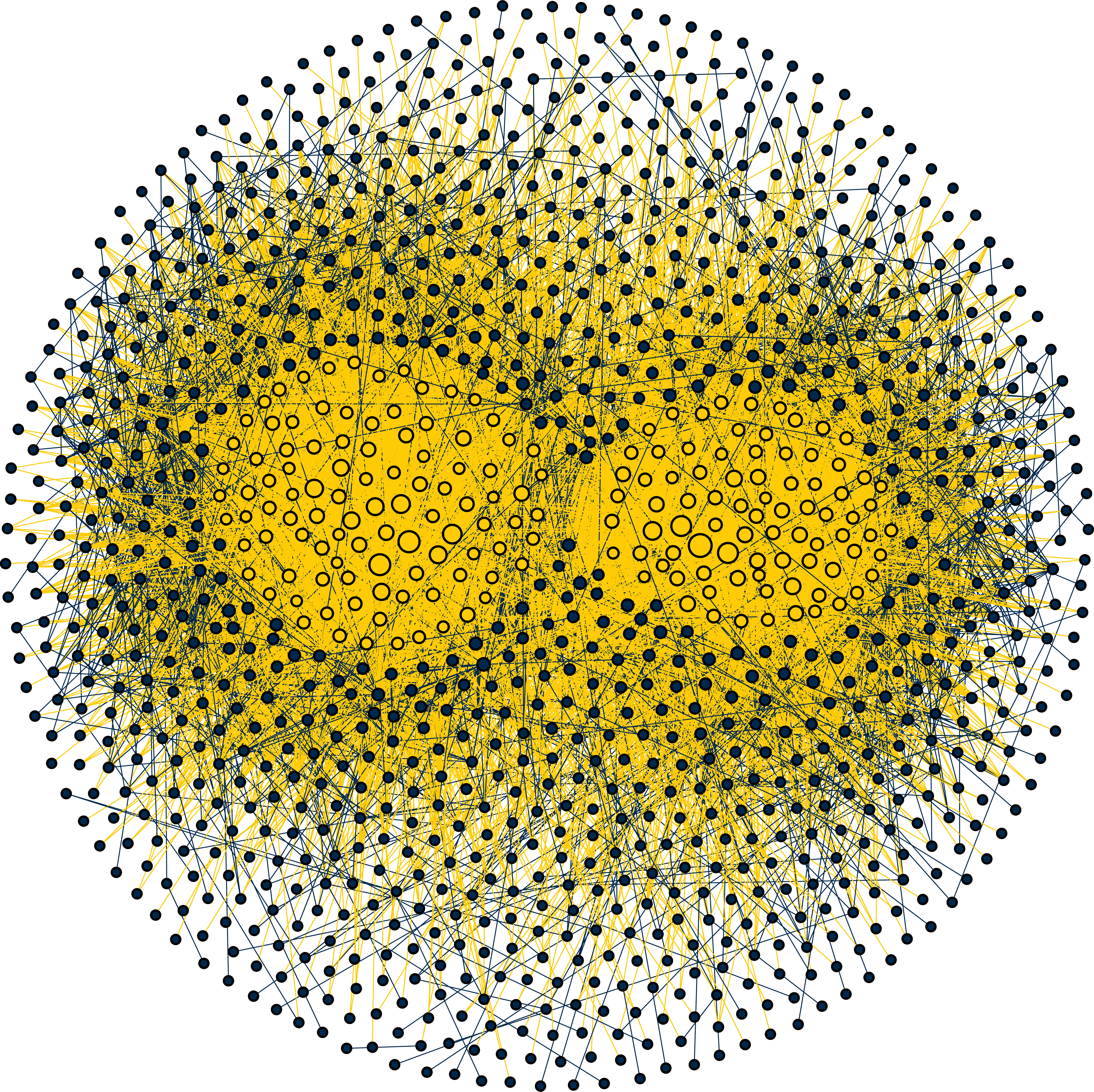}
      \caption{A network of hyperlinks among a set of US political blogs~\cite{AG05}}
      \label{fig:polblogs_2cp}
  \end{subfigure}
  \caption{Two types of two-group core-periphery structure distinguished by the hierarchical model.  Panels (a) and~(b) have a core that is densely connected within itself but only sparsely connected to the periphery.  This is the traditional definition of core-periphery structure.  Panels (c) and~(d) on the other hand show a kind of ``inside-out'' structure in which the core is strongly connected within itself and strongly connected to the periphery.}
\label{fig:inside_out}
\end{figure*}

\section{Example applications}
In this section we give example applications of our methods to a selection of real-world networks, revealing a range of behaviors and structures of interest in the core-periphery divisions of these systems.

\subsection{Traditional two-group core-periphery structure}
For our first set of examples, we perform calculations in which the number of groups is fixed at $k=2$, which corresponds to the traditional two-group core-periphery structure, as studied by many previous authors.  Figure~\ref{fig:inside_out} shows examples of such structure found in four different networks.  For each network the structure shown is the highest-probability structure found during a single run of our algorithm with $10^9$ Monte Carlo steps and in each case the algorithm finds clear core-periphery divisions, as highlighted by the colors.

A number of interesting features emerge in these examples.  First, we note that, as our model is defined, it is arguably the \emph{edges} that belong to groups, not the nodes.  As we have said, a node can belong to any number of groups, but an edge only belongs to one: the properties of each edge are determined solely by the highest-numbered group to which its two nodes both belong and in this sense the edge belongs to this group only.  In Fig.~\ref{fig:inside_out} we have colored the edges according to the group they belong to and this provides a clear and useful visualization.  (The same trick will also be useful in Section~\ref{sec:multiple} when we study divisions with larger numbers of groups.)

All the images in Fig.~\ref{fig:inside_out} use the same color scheme: group~0 is in yellow and group~1 is in blue.  The figure reveals that there are two distinctly different types of core-periphery structure, one where the core is group~0 and one where it is group~1.  Recall that the probability~$\omega_r$ of connection between two nodes depends on their highest common group~$r$, meaning in this case that edges between nodes that are both in group~1 have probability~$\omega_1$ while all others have probability~$\omega_0$.  With this in mind take a look at the figure.

Figures~\ref{fig:inside_out}(a)~and~(b) show results for a network of airline routes~\cite{CGZ13} and a network of associations among a group of terrorists~\cite{Hayes06} respectively.  In both of these cases the core found in the network is represented by group~1 (in blue) and the periphery by group~0 (in yellow), with $\omega_1>\omega_0$.  This implies that there is a high probability of edges within the core (blue edges) and a lower probability both in the periphery and also between the core and the periphery (yellow edges).

Conversely, in Figs.~\ref{fig:inside_out}(c)~and~(d), which represent the Internet at the autonomous system level~\cite{LA14} and a network of political weblogs~\cite{AG05}, the groups are reversed, with the core being group~0 and the periphery being group~1, and $\omega_1<\omega_0$.  In this ``inside-out'' type of structure there is a high probability of connections both within the core and between the core and periphery (yellow edges), and a lower probability in the periphery (blue edges).

These two types of core-periphery structure represent quite different circumstances.  In the first, the core is isolated from the periphery in the sense that it is densely connected only within itself and sparsely connected to everything else.  In the second, the core is strongly connected everywhere, both to itself and to others and dominates the connectivity of the network.  The latter (``inside-out'') structure is particularly interesting because it deviates from the traditional definition of core-periphery structure as formulated for instance by Borgatti and Everett~\cite{BE99}, who assumed an isolated core.  Our method naturally and automatically distinguishes between the two types of structure.

The two types make some sense in the present case.  For the airline route network, for instance, the core broadly represents airline hubs and the periphery represents regional airports.  One expects strong connections between hubs---almost all pairs of hubs have direct flights---but one expects only weak connections to the outlying airports, many of which only fly to a single hub.  Conversely, in the weblog network, for example, the core represents the most influential blogs, ones which most members of the community link to, so we expect connections to be strong not only within the core but also between the core and the periphery.

\begin{figure}
  \includegraphics[width=\columnwidth]{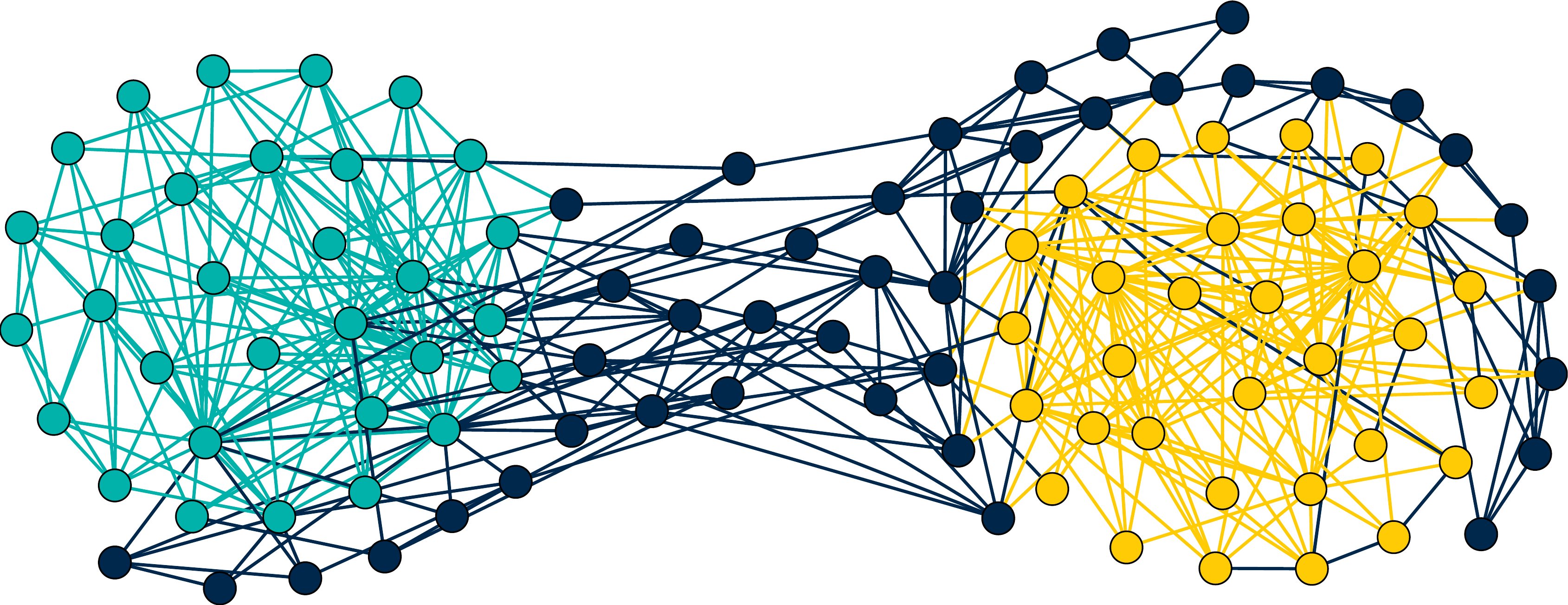}
  \caption{Political books network with a periphery and two cores corresponding to left and right leaning books.}
  \label{fig:polbook}
\end{figure}

\subsection{Structure with an arbitrary number of groups}
\label{sec:multiple}
Now let us look at what happens when we allow the number of groups to vary, taking whatever value is necessary to best fit the structure of the network.  Here again we find some interesting features.  As a first example, Fig.~\ref{fig:polbook} shows a copurchasing network of books.  The nodes in this network represent 105 popular books on US politics and the edges represent frequent copurchase on Amazon.com, i.e.,~purchase by the same buyers.  This network, which has been studied previously by a number of authors~\cite{Newman06b,peixoto21}, is known to show clear community structure in which the network divides into communities of left- and right-leaning books.  Our core-periphery analysis, as indicated by the colors in the figure, finds three groups: two cores and a single periphery.  The two cores correspond to the innermost members of the left- and right-leaning communities while the periphery captures the remainder of the network.  Thus, the algorithm has found the political divide between left and right but also finds a large group of peripheral books that, at least in this analysis, are well represented as a homogeneous mass, suggesting that they are not strongly connected to either side of the political aisle.

\begin{figure}
  \includegraphics[width=\columnwidth]{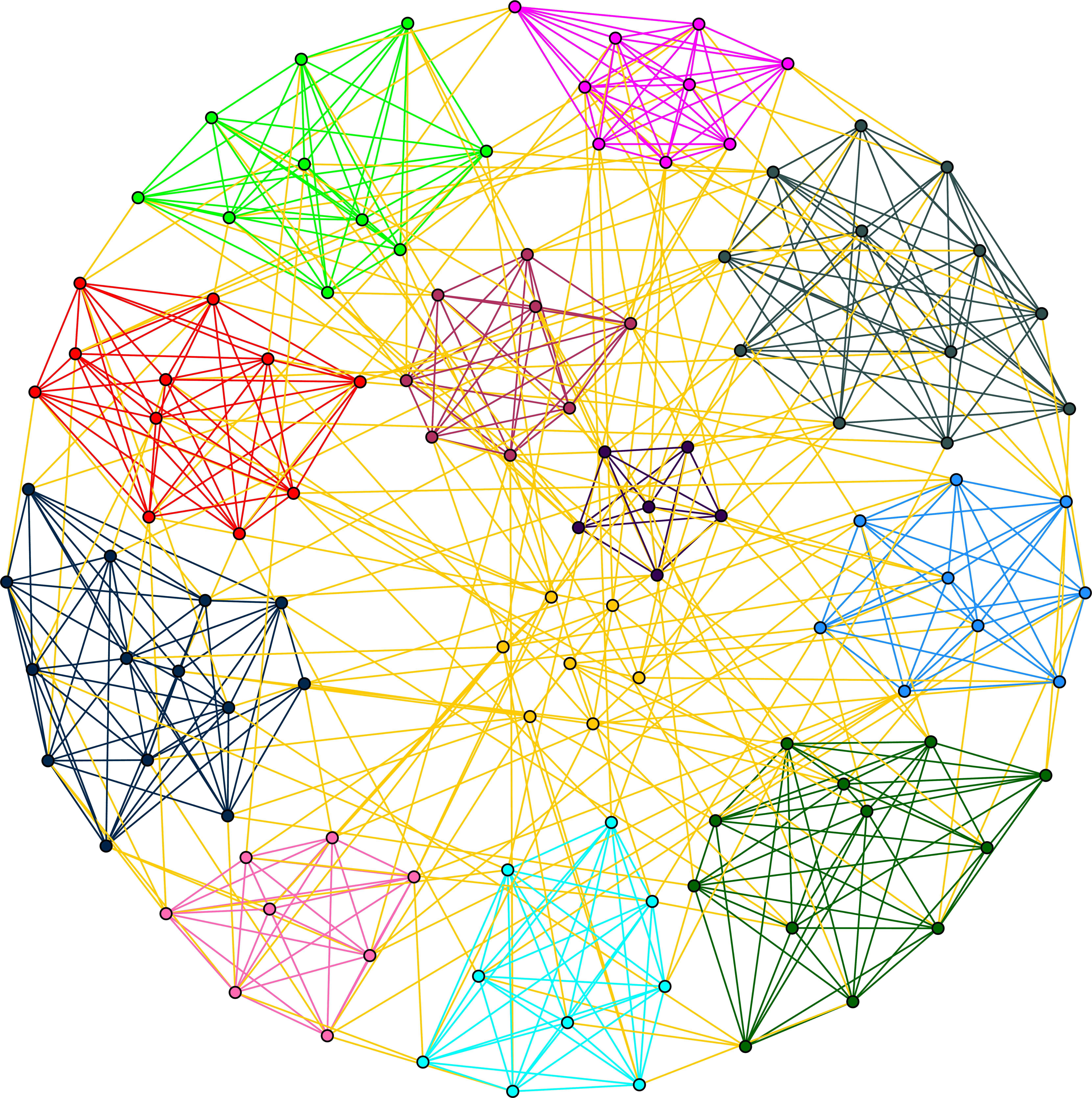}
  \caption{American Football network where each of the cliques are connected to each other via the periphery.}
  \label{fig:amfoot}
\end{figure}

Figure~\ref{fig:amfoot} shows a similar finding for another well studied example of community structure, a network of competition between US college teams in the sport of American football~\cite{GN02}.  College football teams are divided into a number of groups or ``conferences,'' and most games are played between teams in the same conference, so the network of games played, as analyzed here, has strong community structure which can easily be discovered with a range of community detection algorithms.  Again, however, our core-periphery analysis returns a more subtle picture, as shown in the figure.  Our algorithm finds a separate core for each conference, accurately dividing most teams into the 11 conferences in the network.  A small number of teams---many of them independents who belong to no conference---are not assigned to any core, and all inter-conference games are assigned to the periphery.  This makes good sense: it tells us that the conferences constitute a clear set of separate groups in the network, while inter-conference play and non-conference teams constitute a single periphery.  This is an accurate description of the network and a more economical one than the standard community structure division, as found for instance using the stochastic block model, which also assigns a separate community for each conference but in addition assigns a separate probability for inter-conference play between every single pair of conferences, rather than recognizing that a single periphery is an adequate and more parsimonious description.

\begin{figure}
  \begin{subfigure}{\columnwidth}
      \includegraphics[width=0.8\textwidth]{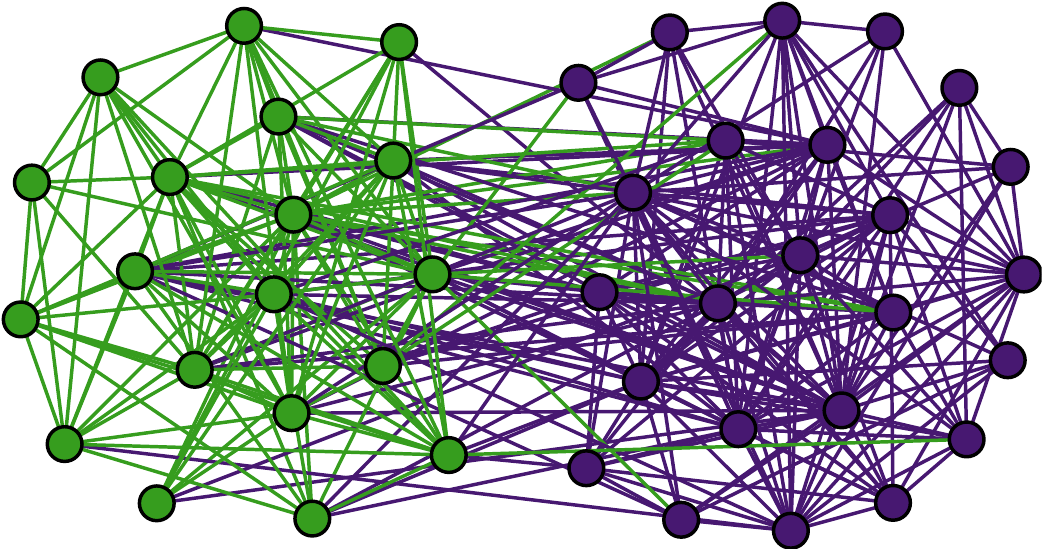}
      \caption{Traditional community structure}
      \label{fig:ws_comm}
  \end{subfigure}
  \hfill
  \begin{subfigure}{\columnwidth}
      \includegraphics[width=0.8\textwidth]{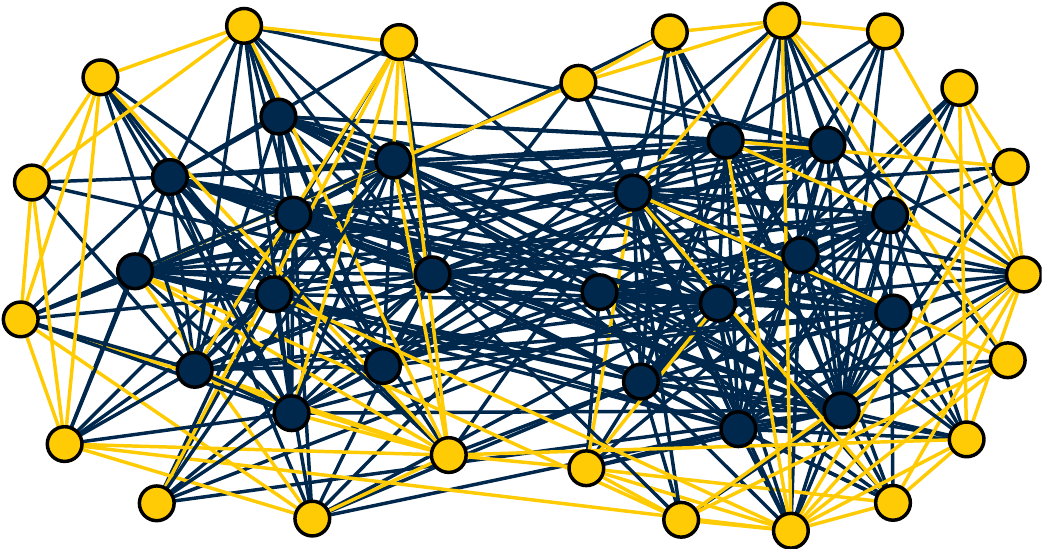}
      \caption{Core-periphery structure}
      \label{fig:ws_cp}
  \end{subfigure}
  \caption{Structure found in the network of windsurfers.}
\label{fig:ws}
\end{figure}

In these last two examples our algorithm has found a hybrid of core-periphery structure and community structure.  While this is illuminating for these particular examples, it is important to realize that this is not inevitable, and that the algorithm will return other structures where appropriate.  Figure~\ref{fig:ws} shows an example.  The network in this figure is a famous one from the social networks literature, a network of interactions observed by Freeman~\cite{FFM88} between a group of people windsurfing off the California coast in 1986.  This network is known to have a clear two-group community structure which is easily found by community detection---see Fig.~\ref{fig:ws_comm}.  When analyzed using the methods of this paper we also find two groups, but they are not the same: now we find core and periphery but no clear division between the communities, suggesting that connections within the core may be just as important as divisions between the two communities.

\section{Conclusions}
In this paper we have proposed a hierarchical model of core-periphery structure in networks and a Monte Carlo scheme for fitting it to observed network data.  Applying these methods to a variety of real-world networks we find a number of interesting patterns.  The method is able to capture traditional two-group core-periphery structure consisting of a dense core weakly connected to a sparse periphery.  In some networks, however, we find that a better fit is given by a novel ``inside-out'' structure in which the core is connected strongly both within itself and to the periphery.  Various networks are better represented by one or other of the two types of structure and the distinction between the two could offer a more nuanced view of structure and function in these networks.

We have also investigated cases where there are more than two groups in the network, generalizing the traditional core-periphery structure (as other authors have also done).  For this we use a Monte Carlo scheme that allows the number of groups to vary freely, automatically choosing the number that best fits the network in question.  In some cases, we find a structure akin to a hybrid between core-periphery structure and community structure in which there is a separate core in each of several communities plus a single periphery surrounding all of them.  In other cases, we find pure core-periphery structure without any communities.

There are a number of possible directions for further research using these methods.  First, we have looked here at only the highest probability structures found by our algorithms but in principle the algorithms return a complete sample of high-probability structures drawn from the posterior distribution of the model and it would be interesting to study the range of structures within such a sample.  Are they all closely similar, so that a single consensus structure can well represent them all, or is there significant variation between structures, and if so of what kind?  Second, one could examine generalizations of the method to broader classes of networks, such as directed and weighted networks and multiplex networks.  Another interesting question is whether there exists a natural ``degree-corrected'' version of the model akin to the degree-corrected stochastic block model of~\cite{KN11a}.  The model proposed here is not degree corrected, which could cause issues with networks that have a very broad degree distribution.  These questions, however, we leave for future work.

\begin{acknowledgments}
The authors thank the Simons Institute for the Theory of Computing at the University of California, Berkeley for hospitality while part of this work was conducted.  This work was funded in part by the US National Science Foundation under grant DMS--2005899.  Computer code implementing the methods of this paper is available at \url{https://github.com/apolanco115/hcp}.
\end{acknowledgments}

\bibliographystyle{numeric}
\bibliography{journals,references}

\end{document}